\documentclass[9pt,twocolumn,twoside]{pnas-new}

\templatetype{pnasresearcharticle} % Choose template 
\usepackage{siunitx}
\usepackage{setspace}
\usepackage{textgreek}

\title{Self-propulsion of inverse Leidenfrost drops on a cryogenic bath}

\author[a,1]{Ana\"\i s Gauthier}
\author[a,b]{Christian Diddens}
\author[c]{R\'emi Proville}
\author[a,d]{Detlef Lohse}
\author[a]{Devaraj van der Meer}

\affil[a]{Physics of Fluids group and Max Plank Center Twente. Mesa + Institute and Faculty of Science and Technology, J.M. Burgers Centre for Fluid Dynamics and Max Plank Center Twente for Complex Fluid Dynamics. University of Twente, P.O. Box 217 7500 AE Enschede, The Netherlands}
\affil[b]{Department of Mechanical Engineering, Eindhoven University of Technology, P.O. Box 513, 5600 MB Eindhoven, The Netherlands}
\affil[c]{INSERM, Neurocentre Magendie, Physiopathologie de la plasticit\'e neuronale, U1215, 33077, Bordeaux, cedex, France}
\affil[d]{Max Planck Institute for Dynamics and Self-Organization, 37077 G\"ottingen, Germany}

\leadauthor{Gauthier} 

\significancestatement{An inverse Leidenfrost state can happen when ambient temperature drops are deposited on liquid nitrogen, as the bath evaporation generates a vapor film that maintains drops in levitation. Contrary to what is seen on solids, drops levitating on a cryogenic liquid exhibit counter-intuitive dynamics: they are spontaneously self-propelled (and glide in straight lines) and keep levitating even after cooling down to a temperature equal to that of the bath. This spontaneous self-propulsion in a cryogenic environment - that last for tens of minutes - can be seen as an efficient way to freeze and further transport biological materials (such as cells or proteins) or chemicals without contamination or risk of heat degradation.}

\authorcontributions{A.G., D. vdM. and D.L. designed research, A.G. performed experiments, A.G. and R.P. analysed data, C.D. did the simulations and A.G. led the writing.}
\authordeclaration{The authors declare no conflict of interest.}
\correspondingauthor{\textsuperscript{1}To whom correspondence should be addressed. E-mail: a.e.gauthier@utwente.nl}


\begin{abstract}
When deposited on a hot bath, volatile drops are observed to stay in levitation: the so-called Leidenfrost effect. Here, we discuss drop dynamics in an inverse Leidenfrost situation where room-temperature drops are deposited on a liquid nitrogen pool, and levitate on a vapor film generated by evaporation of the bath. In the seconds following deposition, we observe that the droplets start to glide on the bath along a straight path, only disrupted by elastic bouncing close to the edges of the container. Initially at rest, these self-propelled drops accelerate within a few seconds and reach velocities on the order of a few cm/s before slowing down on a longer time scale. They remain self-propelled as long as they are sitting on the bath, even after freezing and cooling down to liquid nitrogen temperature. We experimentally investigate the parameters that affect liquid motion, and propose a model, based on the experimentally and numerically observed (stable) symmetry breaking within the vapor film that supports the drop. When also the film thickness and the cooling dynamics of the drops are modeled, the variations of the drop velocities can be accurately reproduced.
\end{abstract}

\begin{document}

% Optional adjustment to line up main text (after abstract) of first page with line numbers, when using both lineno and twocolumn options.
% You should only change this length when you've finalised the article contents.
%\verticaladjustment{-10pt}

\maketitle
\thispagestyle{firststyle}
\ifthenelse{\boolean{shortarticle}}{\ifthenelse{\boolean{singlecolumn}}{\abscontentformatted}{\abscontent}}{}

% If your first paragraph (i.e. with the \dropcap) contains a list environment (quote, quotation, theorem, definition, enumerate, itemize...), the line after the list may have some extra indentation. If this is the case, add \parshape=0 to the end of the list environment.

%\doublespacing

%%%%%%%%%%%%%%%%%INTRODUCTION%%%%%%%%%%%%%%%%%%%%%

\dropcap{W}hen deposited on a hot solid, volatile drops can levitate over a cushion of their own vapor - a phenomenon extensively described by J.G. Leidenfrost \cite{Leidenfrost:1756} in the 18th century. Being insulated from the substrate by a vapor layer, the Leidenfrost drops have a lifetime of the order of a few minutes \cite{Biance:2003}. Moreover, in absence of friction, they do not only glide at the slightest inclination but also bounce \cite{Tran:2012}, jump \cite{Celestini:2012} or oscillate \cite{Brunet:2011}, rich dynamics {\cite{Quere:2013}} that make the control of such drops a problem. On solid substrates, addition of a well-chosen texture can efficiently guide drops, as first demonstrated by Linke \textit{et al.} \cite{Linke:2006}: asymmetric textures can redirect the vapor flow below the liquid \cite{Dupeux:2011}, which generates self-propulsion. This is used to efficiently guide or even entrap levitating drops \cite{Cousins:2012, Marin:2012, Soto:2016} or solids \cite{Hashmi:2012}. However, controlling drop motion seems more complex on deformable substrates such as liquid baths, where Leidenfrost levitation also occurs \cite{Snezhko:2008, Kim:2011, Maquet:2016, Adda-Bedia:2016, Janssens:2017}. The liquid surface, resisting the weight of the drops is notably deformed  \cite{Vella:2015, Wong:2017}, but this does not impact drop mobility, as there is no contact drag \cite{Vakarelski:2011, LeMerrer:2011}. The suspended drops were observed to sometimes glide for tens of seconds \cite{Song:2010, Snezhko:2008, Adda-Bedia:2016, Janssens:2017, Feng:2018}, and have to be trapped to perform some measurements \cite{Maquet:2016}. 

\parshape=0
In this paper, we consider the dynamics of ethanol or silicone oil droplets deposited on a liquid nitrogen bath, in an "inverse" Leidenfrost scenario \cite{Hall:1969} where vapor generated \textit{ by the bath} maintains drops above the pool. We show that, contrary to what is seen on solid substrates, a spontaneous symmetry breaking occurs that leads to a self-propelling state - a phenomenon that we investigate experimentally. Using simulations, we demonstrate that the movements arise from a difference in the film thickness between the front and the back of the drop, which we use to model the gliding dynamics.

%%%%%%%%%%%%%%%%EXPERIMENT%%%%%%%%%%%%%%%%%%%%%%

\subsection*{Experiment}

\begin{figure*} [h!]
\centering
\includegraphics[scale=0.47]{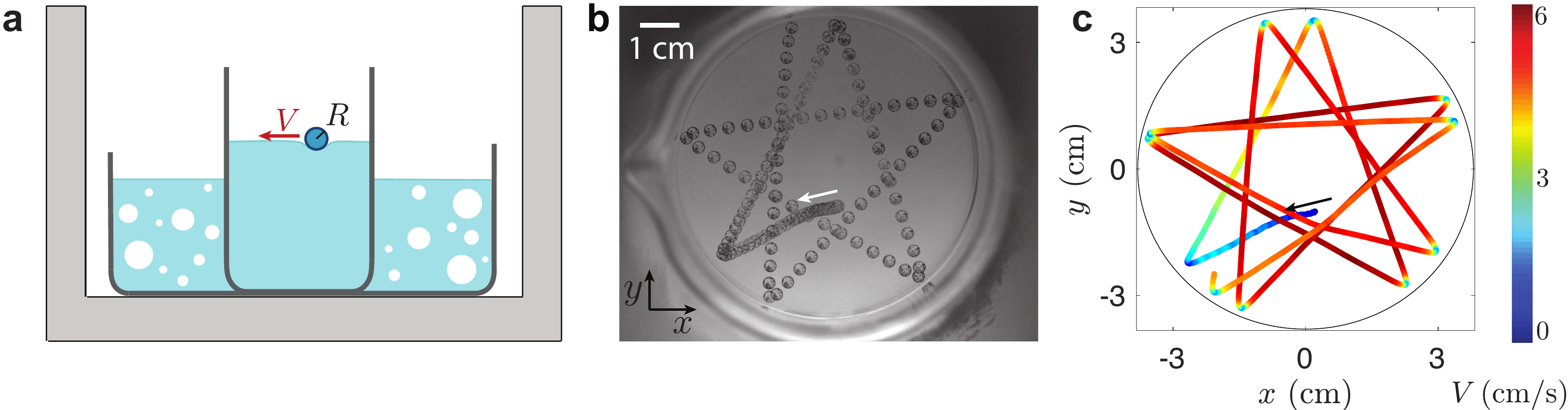} 
\caption{\label{figure1} \textbf{a.} Schematic of the experimental setup: a drop with radius $R$ is deposited on a liquid nitrogen bath. To avoid ebullition of the central bath, it is placed in a styrofoam box, and maintained at the center of a sacrificial bath whose evaporation maintains a nitrogen atmosphere in the box. Drop trajectory and velocity $V$ are recorded from the top. \textbf{b}. Chronophotography of the successive positions (separated by 80 ms) of an ethanol drop ($R$ = 1.5 mm) seeded with particles. The white arrow indicates the initial position and movement of ethanol. \textbf{c}. Trajectory of the center of mass of the same drop in the x-y plane in a longer time interval. The color indicates the drop velocity, varying from $V = 0$ (dark blue) to $V = 6$ cm/s (dark red).}
\end{figure*}

Liquid nitrogen is a cryogenic liquid with boiling temperature of -196$^\circ$C and low latent heat of vaporization $L_v = 2\, \text{x}\, 10^5$ J/kg. Its evaporations is fast enough so that, when a drop at ambient temperature approaches a nitrogen bath, the generated vapor cushion can maintain the drop in the Leidenfrost state \cite{Adda-Bedia:2016, Song:2010}.  As opposed to more usual Leidenfrost situations \cite{Leidenfrost:1756, Biance:2003, Tran:2012} where vapor is produced by the levitating objects, here vapor comes from the bath so that the drops keep a constant radius $R$ over time. However, the drops continuously cool down (below their freezing point), until their temperature reaches that of the bath, which sometimes causes their sinking \cite{Adda-Bedia:2016, Song:2010}. To avoid ebullition within the pool, we followed Adda-Bedia \textit{et al.} \cite{Adda-Bedia:2016} by placing the central bath (with diameter $D$ = 7.6 cm) at the center of a sacrificial bath of liquid nitrogen, itself inside an homemade polystyrene cryostat. As schematized in Figure \ref{figure1}a, the sacrificial bath is continuously boiling, which maintains a nitrogen atmosphere in the box. The residual evaporation of the central bath (at approximately 0.1 L/h, due to radiative heat exchanges at the top) does not disturb the liquid surface that remains perfectly still. Drops of ethanol (density $\rho$ = 789 kg/m$^3$, specific heat $c_p$ = 2400 J/kg/K at 20$^\circ$C) or silicone oil ($\rho$ = 930 kg/m$^3$, $c_p$ = 1600 J/kg/K) with radii $R$ ranging from 0.65 mm to 1.8 mm are formed from calibrated needles and released $\simeq$ 1 cm above the bath surface. The chosen liquids have low freezing temperatures (<  -100$^\circ$C), which limits their freezing in the needles. Moreover, such drops keep a smooth spherical shape when they freeze , which does not always happen for water drops \cite{Wilderman:2017}. Once released, the drops, denser than liquid nitrogen (density $\rho_N$ = 808 kg/m$^3$), initially sink; but nitrogen evaporation generates a buoyant force that almost immediately pushes them back to the surface, where they remain \cite{Song:2010}. Drop trajectory and velocity $V$ are recorded from the top at typically 125 fps, and the origin of time $t$ is chosen as soon as Leidenfrost levitation happens.

\smallskip

Figure \ref{figure1}b shows the first 15 seconds of motion of an ethanol drop with radius $R$ = 1.5 mm seeded with particles - two successive images are separated by 80 ms. The white arrow indicates the initial position and direction of the drop. Ethanol, initially at rest, slowly accelerates and starts hovering on the bath in straight lines. This regular movement is only disturbed by almost perfect reflections close to the edges of the beaker, producing a remarkable star-shaped trajectory. The droplet is initially subjected to strong internal motion (that can be seen on Movie S1) that vanishes as the liquid cools down and freezes (which happens between the second and the third bouncing) - with no visible impact on its movement. As visible in Movie S1, the surface of the bath remain still as the drop hovers it. The drop velocity $V$, of a few centimeters per second, is small enough not to generate any stationary wake \cite{LeMerrer:2011}.

In Figure \ref{figure1}c, the trajectory of the drop center of mass is plotted: the position (x,y) = (0,0) is the center of the beaker and the black circle corresponds to the edge of the bath. The color code indicates the drop velocity $V$. Initially 0 (dark blue), $V$ increases up to 6 cm/s (dark red) after the 4${^\text{th}}$ bouncing, and then slowly diminishes to reach 4 cm/s after the 13$^{\text{th}}$ bouncing. Interestingly, the propulsion mechanism is not disturbed by the successive rebounds: in the first instants, the drop keeps accelerating even after turning back close to the edge.
The setting in motion of the drops is observed for every liquid tested (ethanol, silicone oil, propanol, butanol, pentanol, water), provided the drops were small and light enough to be supported by the liquid nitrogen bath. Depending on the first incident angle of the drop with the wall, trajectories vary from diagonals (for a perfectly normal incidence) to stars with varying number of branches - as in the SI Appendix, Fig.1 - up to triangles, pentagons and circles (for a tangent impact). Finally, it can be noted that self-propulsion is also seen for millimeter-sized particles (polyethylene spheres), although for a much shorter duration. Similarly to frozen drops, the solid particles do not exhibit any rotational movement while gliding (as in Movie S2).

\begin{figure}[h]
\includegraphics[width=0.49\textwidth]{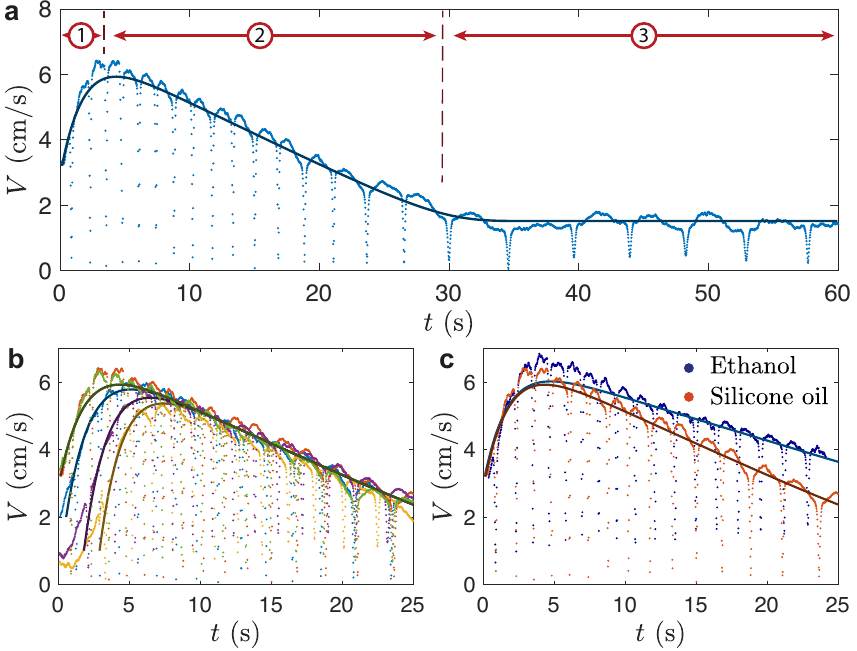}
\caption{\label{figure2} \textbf{Drop velocity} \textbf{a.} Velocity $V$ of a drop of silicone oil ($R$ = 1.4 mm) gliding on a liquid nitrogen bath, as a function of time $t$. The numbers mark the 3 phases of movement: acceleration, deceleration and constant velocity. The corresponding movie is Movie S3. \textbf{b.} Comparison of the velocity $V(t)$ of 5 identical silicone oil drops ($R$ = 1.4 mm) deposited slightly differently on the bath. \textbf{c.} Velocity $V(t)$ of ethanol (blue dots, specific heat $c_p$ = 2400 J/kg/K) and silicone oil drop (orange dots, $c_p$ = 1600 J/kg/K) with similar radius ($R$ = 1.4 mm) and initial velocity. In \textbf{a}, \textbf{b} and \textbf{c}, the darker lines are the numerical solution of Eq. \ref{NewtonsLaw}, with identical prefactors $\alpha$ = $\beta$ = 15, and $\Delta h$ =~1.45\, \textmugreek m}
\end{figure}

The velocity dynamics $V(t)$ is even more intriguing. Figure \ref{figure2}a shows $V(t)$ for a silicone oil drop ($R$ = 1.4 mm) as it glides on the bath (see also Movie S3). After falling from the needle, the drop sinks and resurfaces with an initial velocity $V(t=0)$ = 3.2 cm/s, and immediately accelerates. The shape of $V(t)$ results from the combination of two effects. First, at each rebound, the velocity $V$ decreases and rises up again to the same value - indicating elastic bouncing. The drop bounces 23 times during its 60 s lifetime: each event can be distinguished individually in Fig. \ref{figure2}a. Second, $V(t)$ exhibits very regular variations on a longer timescale (variations that are quite undisturbed by the repeated bounces) and that we call here the \textit{velocity amplitude} $V_A(t)$. $V_A$ is highlighted by the black line (which is the numerical solution of Eq. \ref{NewtonsLaw}) and can be decomposed in 3 phases, numbered on Fig. 2a:
(1) An \textit{acceleration} phase (for $0 < t < 5$ s) where the drop velocity amplitude increases from $V_A$ = 3.2 cm/s to 6.5 cm/s.
(2) A \textit{deceleration} phase (for 5 s $< t < 30$ s) that lasts 5 times longer than the acceleration, and during which $V_A$ decreases linearly with time.
(3) A \textit{constant velocity} phase (30$\,$s$\,<\,t\,<\,60\,$s) with $V_A \simeq 2$ cm/s. This third phase can sometimes last several minutes, until an outside event (a small movement of the liquid surface or an encounter with a floating ice cristal) makes the drop sink. The levitation time is much longer than the expected Leidenfrost duration, which is of the order of $30\,$s for millimeter-sized drops initially at ambient temperature \cite{Adda-Bedia:2016}. It can also be noticed that drop immersion after more than 30~s hardly generates any boiling (see Movie S4), which indicates that the particle temperature is then close to the vaporization temperature of the bath.

Figure \ref{figure2}b shows 5 velocity plots (colored dots) obtained by repeating the same experiment (silicone oil, $R$ = 1.4 mm) but varying the height at which the drops are deposited. If the velocity amplitudes of drops with identical initial velocity perfectly overlap (as for the green and red curves), varying the initial conditions impacts the acceleration phase. In particular, it can be noted that sometimes (yellow and purple plots), the drops do not accelerate immediately but exhibit an erratic motion at low velocity ($V$ < 1cm/s) for the first seconds before starting to self-propel. Interestingly, this does not impact the second phase of the drop movement (the deceleration) where all $V_A(t)$ plots perfectly overlap.
Finally, Fig. \ref{figure2}c compares the velocity profiles of ethanol and silicone oil with identical radius $R$ = 1.4 mm and initial velocity. Contrary to Fig\,\ref{figure2}b, changing the nature of the liquid affects the deceleration rate, which is significantly (30$\%$) lower for ethanol than for silicone oil. However, varying the drops freezing temperature, or preheating them hardly influences the velocity amplitude (see the SI Appendix, Fig.2). We now aim to understand and model the phenomena at the origin of the rich drop dynamics.

\subsection*{Origin of self-propulsion}  

\begin{figure}[h!]
\includegraphics[width=0.49\textwidth]{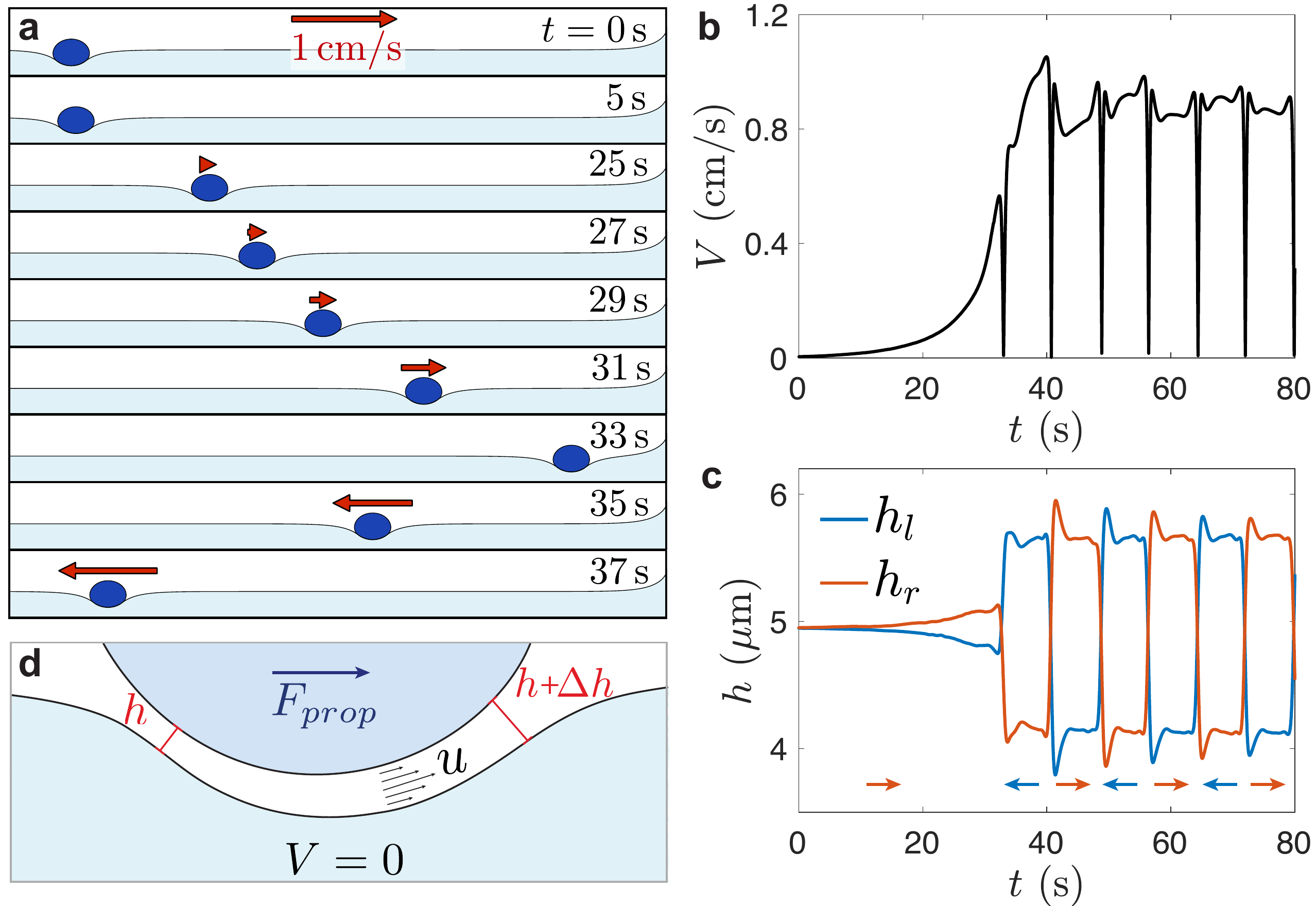}
\caption{\label{figure3} \textbf{a}. Successive images extracted from the 2d-simulations. A drop with viscosity $\eta$ = 16 mPas.s and radius $R$ = 1 mm is deposited on an evaporating bath, and spontaneously self-propels. The corresponding movie is Movie S6. \textbf{b}. Drop velocity $V$ as a function of time $t$. \textbf{c}. Difference in film thickness $h$ between the left ($h_{l}$, in blue) and right part ($h_{r}$, in red) of the drop where the film is thinner. \textbf{d}. Model for the propulsion force: the film has a mean thickness $h \ll R$, and vapor is continuously escaping with a characteristic velocity $u$. The difference $\Delta h \ll h$ in the film thickness changes the vapor distribution, generating a viscous propelling force $F_{prop}$, directed towards the larger opening.} 
\end{figure}

%%Begin Numerical part
A first insight on the cause of self-propulsion is obtained through numerical simulation. Both vapor and liquid flows are calculated in a 2d model system, using a sharp-interface finite element method. A drop is deposited at the center of a liquid nitrogen bath and the initial vapor film is symmetric. As visible in Movie S5, the mesh is made very fine below the drop - to resolve the thin gas film - and coarser outside. For simplicity, thermal effects are neglected and the bath evaporates at a constant rate of 2.15 g/s$^2$. The motion of a drop with radius $R$~=~1~mm and viscosity $\eta$ = 16 mPa.s, as obtained numerically is presented in Fig. \ref{figure3}a: even if no pre-existing asymmetry is imposed, the droplet spontaneously self-propels, as also visible in Movie S6. In Fig \ref{figure3}b, the drop velocity $V$ is plotted as a function of time: $V$ increases to finally reach a constant value $V_A^* = 0.85 \,  \pm \,$0.1\,cm/s. Similarly to what is seen experimentally, the velocity amplitude is not significantly impacted by the repeated drop about-turns close to the edges of the bath. 

Beyond a mere reproduction of the self-propulsion, the simulation gives access to the details of the film thickness and its variation with time, information difficult to obtain experimentally. In figure \ref{figure3}c, the minimum film thicknesses (measured at the neck) on the left side of the drop ($h_l$, in blue) and on the right side ($h_r$, in red) are extracted from the simulation and plotted as a function of time. While initially, $h_l$ and $h_r$ are equal, they spontaneously diverge until a constant asymmetry $\Delta h =  | h_{l} - h_{r} |  \simeq 1.5 \, \pm 0.1 \,$\textmugreek m is reached. Thus, Figure \ref{figure3}c gives essential indications on the origin of self-propulsion. First, comparison with Fig. \ref{figure3}b shows that the appearance of the assymmetry corresponds to the setting in motion of the drop. A geometrical asymmetry would indeed partially redirect the flow of vapor towards the larger opening, and thus generate a propelling force. In addition, the film is systematically thicker at the front ($h_{r}$ when the drop moves to the right, $h_{l}$ when it moves to the left). The drop follows the preferential direction of motion of the vapor, which indicates that the mechanism that causes self-propulsion is surely of viscous origin. Finally, it should be noted that the asymmetry spontaneously switches from left to right when the drop gets close to the liquid meniscus at the edge of the bath. While the direction of $\Delta h$ changes, its amplitude is not impacted: the same asymmetric state consistently reappears. This strongly suggests that the symmetric film thickness is metastable, and that self-propulsion is generated by a spontaneous and constant symmetry breaking within the vapor film.

\subsection*{Model} The main result of the simulation is now used to model drop dynamics. As observed in the numerics, we assume a constant asymmetry with amplitude $\Delta h$ in the film thickness $h$ (with $\Delta h \ll h$) between the front and the back of the drop. As illustrated in Fig. \ref{figure3}d, and similarly to what is seen on textured solids \cite{Linke:2006, Dupeux:2013}, the asymmetry partially redirects the flow of vapor, which enable motion. The difference of viscous stresses between the front and the back generates a propelling force $F_{prop}$ which can be estimated: $F_{prop}$ is a fraction $\Delta h/h$ of the total viscous force exerted on the bottom of the drop, varying as $\eta_v \frac{u}{h} \, R^2$ (with $\eta_v$ the viscosity of nitrogen vapor and $u$ the typical velocity of the Poiseuille flow within the film), which gives $F_{prop} \sim \frac{\Delta h}{h} \; \eta_v \frac{u}{h} R^2$. This expression is simplified using lubrication theory: the pressure drop  $\Delta p$ within the vapor film scales as $\Delta p \sim \eta_v \frac{u}{h^2} R$ and the overpressure in the film sustains the drop, which implies $\Delta p \sim \rho g R$ for drops smaller than the capillary length \cite{Biance:2003, Adda-Bedia:2016}. These three expressions combined give the following propelling force:

\begin{eqnarray}
F_{prop} \sim \rho g R^2 \, \Delta h,
\label{eqFprop}
\end{eqnarray}

\noindent
which is similar to what is observed for uneven Leidenfrost solids \cite{Dupeux:2013}. $F_{prop}$ dominates at the first instants of motion, during acceleration; but as the drop velocity increases, the friction force $F_f$ gains importance. Our hypothesis is that its dominant contribution also comes from the film: while gliding (with velocity $V$) the drop entrains vapor and its movement creates a secondary Couette flow within the film, with a mean velocity $\propto V$. This generates a viscous friction force that can be written as: 

\begin{eqnarray}
F_{f} \sim \eta_v \frac{V}{h} \, R^2.
\label{eqFriction}
\end{eqnarray}

\noindent The full calculation, (in a simplified situation) confirms this argument and is given in the SI Appendix. The friction force (inversely proportional to the film thickness $h$) has the unusual property of increasing with time. Indeed, as the drop cools down, less and less nitrogen vapor is produced and the film thins out. To fully determine $F_f(t)$, we then need to model $h(t)$. Calculations of the film thickness have been done with a different purpose for levitating drops on solids \cite{Biance:2003}, or on hot baths \cite{Maquet:2016}, and in the inverse Leidenfrost state \cite{Adda-Bedia:2016}: we follow a similar line of arguments here.

For drops in an inverse Leidenfrost scenario, $h(t)$ arises from two simultaneous processes: (i) vapor production and escape and (ii) drop cooling dynamics. We give here the main physical ingredients (a detailed calculation can be found in the SI Appendix). (i) Due to the temperature difference $\Delta T$ between the drop and the bath, heat diffuses through the film and vaporises liquid nitrogen. The escaping vapor is then confined below the drop, and lubrication generates an overpressure that sustains the drop. For any given $\Delta T$, these two elements give the following scaling law for $h$ : 

\begin{eqnarray}
 h \sim \left( \frac{\eta_v \lambda \Delta T R}{\rho g \rho_v L_v} \right )^{1/4} 
 \label{eqH(DeltaT)}
 \end{eqnarray}
\noindent where $\eta_v$, $\rho_v$ and $\lambda$ respectively denote the viscosity, density and conductivity of the vapor, $L_v$ the latent heat of vaporisation of liquid nitrogen, $\rho$ drop density and $g$ gravity. (ii) Simultaneously, due to heat diffusion through the film, the drop cools down. The rate of decrease of the drop internal energy $\rho R^3 c_p \frac{d \Delta T}{dt}$ (with $c_p$ the drop specific heat) is equal to the rate at which energy diffuses through the vapor film $ \lambda \Delta T/h R^2$. Combined with Eq. \ref{eqH(DeltaT)}, integration of this differential equation finally gives $h(t)$, which is found to decrease linearly with time:

\begin{eqnarray}
 h  \sim  h_0 \left(1-t/\tau \right) \; \text{with } \; 
h_0 & \sim &\left( \frac{\eta_v \lambda \Delta T_0 R}{\rho g \rho_v L_v} \right )^{1/4} \label{eqH} \\
\text{and } \; \tau & \sim & \frac{4 \rho R c_p h_0}{\lambda}.  \nonumber 
\end{eqnarray}

\noindent
$h_0$ is the initial film thickness: for millimeter-sized drops and $\Delta T_0 \simeq 200^\circ$C, $h_0 \simeq 50 \,$\textmugreek m. This is in good agreement with measurements done on solid substrates \cite{Biance:2003} or with the results of numerical calculations \cite{Maquet:2016} for drops on a bath. The characteristic time $\tau$ arises from the drop cooling dynamics: it is the time needed for drops to cool down from ambient temperature to liquid nitrogen temperature. For millimeter-sized drops, this time is of the order of 20 s. 
\vspace{0.5 cm}

Using Eq.s \ref{eqFprop}, \ref{eqFriction} and \ref{eqH} we can finally model the dynamics of the droplet. Writing $m$ for the drop mass, Newton's second law gives the following differential equation for the velocity amplitude $V_A(t)$:

\begin{eqnarray}
m \frac{dV_A}{dt} = - \alpha \; \eta_v \frac{V_A}{h(t)} R^2 + \beta \; \rho g R^2 \Delta h 
\label{NewtonsLaw}
\end{eqnarray}

\noindent 
with $\alpha$ and $\beta$ numerical coefficients arising from geometrical factors (respectively for $F_{f}$ and $F_{prop}$) which are not considered in scaling laws. Since $F_{f}$ and $F_{prop}$ both originate from the vapor flow within the film, we can assume that $\alpha$ and $\beta$ are close. Thus, in the rest of the discussion, we will consider $\alpha \simeq \beta$, and the two fitting parameters in Eq. \ref{NewtonsLaw} are $\alpha$ and $\Delta h$. 

\noindent The temporal dependence of the velocity amplitudes $V_A(t)$ is simply deduced from Eq. \ref{NewtonsLaw} using a separation of times scales. On the one hand, $h(t)$ varies in a time $\tau \simeq$ 20 s while, on the other hand, the characteristic time of the acceleration phase is $\tau_{acc} \sim \rho R h/\eta_v \simeq$ 1 s. Therefore, during the acceleration phase the film thickness $h(t)$ remains roughly constant, and Eq. \ref{NewtonsLaw} can be approximated by a first-order linear differential equation. Denoting $V_{A0}$ the initial drop velocity, $V_A(t)$ increases exponentially, with $V_A(t) = V_{A0} + (V_A^*-V_{A0}) \big(1- \exp(-t/\tau_{acc}) \big)$, until the drop reaches its terminal velocity $V_A^*$, obtained by equalizing the propelling and friction forces (Eq. \ref{eqFprop} and \ref{eqFriction}): 

\begin{eqnarray}
V_A^*(t) =  \frac{\rho g}{\eta_v} \, \Delta h \, h(t).
\label{eqTermV}
\end{eqnarray}

\noindent
On a longer time scale, film thinning affects $V_A^*(t)$ that decreases linearly with time (as $h$ does, from Eq. \ref{eqH}) in a characteristic time $\tau$. This model nicely reproduces the first 2 phases of the drop movement, as seen in figures \ref{figure2}b and c: the darker lines are the numerical solution of Eq. \ref{NewtonsLaw}, with the same fitting parameters $\alpha$ = 15 and $\Delta h$ = 1.45 \textmugreek m. In figure \ref{figure2}b, the collapse of the $V_A(t)$ plots during the deceleration phase is due to the drops reaching their terminal velocity, identical for all five experiments. However, the nature of the liquid affects the deceleration rate, as illustrated in Figure \ref{figure2}c. The difference in deceleration rate between ethanol and silicone oil is mainly due to a difference in the liquid specific heats $c_p$. Ethanol drops with $c_p$ = 2400 J/kg cool down in $\tau \simeq$ 43\,s where silicone oil drops (with $c_p$ = 1600 J/kg) cool down faster ($\tau \simeq$ 30\,s); which directly impacts the slope of $V_A^*(t)$.

\begin{figure}[h]
\includegraphics[width=0.49\textwidth]{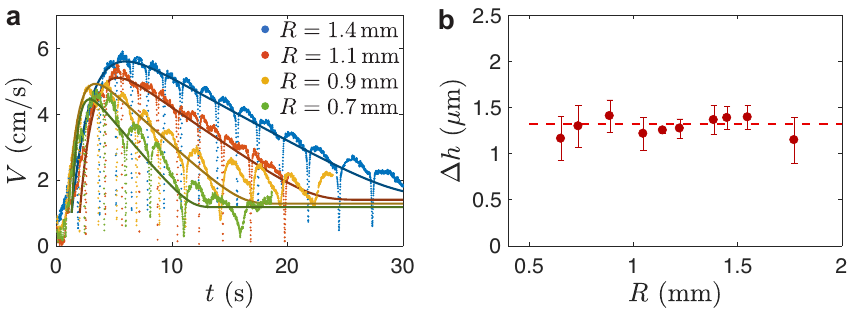}
\caption{\label{figure4} \textbf{a.} Velocity $V(t)$ of silicon oil drops with varying drop radius $R$. The darker lines are the solutions of Eq. \ref{NewtonsLaw} for each drop size. \textbf{b.} Amplitude of the asymmetry $\Delta h$ deduced from the best fit of the drop deceleration.}
\end{figure}

In addition, the amplitude of the asymmetry $\Delta h$ causing self-propulsion can be deduced from the velocity amplitude dynamics. Figure \ref{figure4}a shows the best fit of $V_A(t)$ obtained for varying drop radii $R$: smaller drops have lower internal energy and cool down faster, which is nicely reproduced by Eq \ref{NewtonsLaw}. By repeating systematically this experiment (with drop radius $R$ varied between 0.64 and 1.8 mm) we plotted (in Fig. \ref{figure4}b) the value of $\Delta h$ giving the best fit as a function of $R$ . More specifically, we considered the second phase of drop dynamics, where drops decelerate at a constant rate $a$ (phase 2 in Fig \ref{figure2}a). The experimental measurement of $a$ gives $\Delta h$, that is expected to vary proportionaly to $a \frac{\eta_v \tau}{\rho g h_0}$, as calculated from Eq. \ref{eqTermV}. In Figure \ref{figure3}c, $\Delta h$ is found to be of the order of 1 \textmugreek m, which is consistent with our initial hypothesis of a small film deformation ($\Delta h \ll h \simeq 50 \,$\textmugreek m) and with the result of the numerical simulation. Remarkably, $\Delta h$ remains constant over the range of drop radii we tested (0.64 mm < $R$ < 1.8 mm). This is quite different from what is seen for self-propelled uneven solids \cite{Dupeux:2013} where $\Delta h \propto \sqrt{R}$.

\smallskip

Incidentally, Eq. \ref{NewtonsLaw} explains why the drop does not rotate in the stable asymmetric state: reaching the terminal velocity, the propelling and friction forces balance, which implies - for an approximately spherical drop - that also the net torque on the drop balances, consistent with our observations.

\subsection*{Self-propulsion of pool liquid and frozen drops}
\begin{figure}[h]
\includegraphics[width=0.49\textwidth]{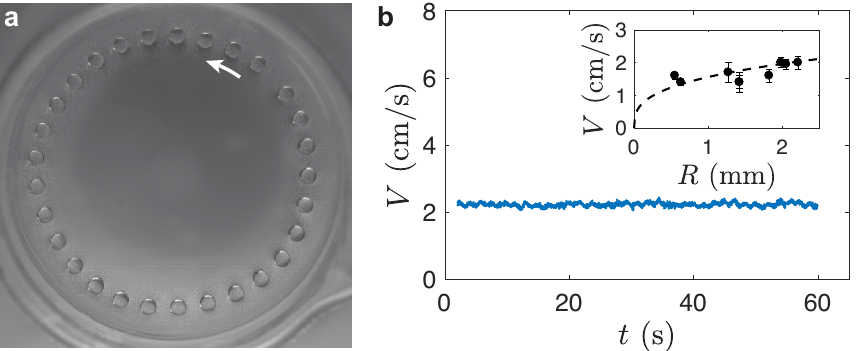}
\caption{\label{figure5} \textbf{a.} Successive positions of a drop of liquid nitrogen with radius $R$ = 1.8 mm deposited close to the edge of a liquid nitrogen bath. The images are separated by 250 ms. The arrow indicates the direction of movement of the drop. Corresponding movie is Movie S7. \textbf{b.} Velocity $V(t)$ of the liquid nitrogen drop. $V$ stays constant, with $V = V_A^* = 2.2 \pm 0.2$ cm/s. In the inset, $V$ is plotted as a function of the drop radius $R$. The dotted line is a fit with $V_A^* = 1.7 \, \frac{\rho g}{\eta_v} \Delta h \, h_{N_2}$, with $h_{N_2}$ as defined in Eq. \ref{eqH1}.}
\end{figure}

Interestingly, liquid nitrogen drops deposited delicately on the liquid nitrogen bath can also levitate (without coalescing) for long periods of time, even if their temperature is the same as that of the bath. Similarly to hot drops, these cryogenic drops are self-propelled: Figure \ref{figure5}a shows successive positions (separated by 250 ms) of a liquid nitrogen drop with radius $R = 1.8$ mm 10 minutes after being deposited. As also seen in Movie S7, the drop has a regular circular trajectory. Such a trajectory is generated because the drop, released close to the edge of the beaker, is initially propelled almost tangentially to the wall. In Fig. \ref{figure5}b the drop velocity $V(t)$ is observed to remain constant, with $V(t) = V_A^* = 2.2 \, \pm$ 0.2 cm/s. 

Levitation of such cold objects is made possible by residual bath evaporation, happening despite the presence of an insulating box and a sacrificial bath. A 300 mL beaker with surface 57 cm$^2$ typically empties in 3h, which corresponds to an evaporation rate $\dot{M} \simeq 10^{-5}$ kg/s. This value is in close agreement to what is expected from radiative heat transfert where $\dot{M}\sim \sigma \left (T_{amb}^4 - T_{N}^4 \right) D^2/L_v$ $\simeq 2 \, \text{x}\, 10^{-5}$ kg/s, with $D$ the beaker diameter, $\sigma$ the Stefan-Boltzmann constant, $T_{amb}$ and $T_N$ respectively ambient and liquid nitrogen temperatures. 

The continuous vapor production maintains a constantly renewed vapor film under the liquid nitrogen drops, even if the droplet itself does not transfer heat to the system. Levitating is then the same as floating above a perfectly porous substrate through which gas escapes. The film thickness $h_{N_2}$ generated by the bath evaporation can be estimated in that situation: the vapor flux generated under the surface of a (cold) drop with radius $R$ is $q \sim \frac{\dot{M} R^2}{\rho_v D^2}$. This vapor is redirected within the film, so that $q \sim u R h_{N_2}$, with $u$ the mean velocity of vapor. The pressure in the film $\Delta p$ is estimated from lubrication theory $\Delta p \sim \eta_v R u/h_{N_2}^2$ and as the vapor film sustains the drop $\Delta p \sim \rho g R$. These expressions combined finally give:

\begin{eqnarray}
h_{N_2} \sim \left(\frac{\eta_v \sigma \left (T_{amb}^4 - T_{N}^4 \right)R}{\rho_v \rho g L_v} \right)^{1/3}
\label{eqH1}
\end{eqnarray}

The measured evaporation rates yield $h_{N_2} \simeq$ 10 \textmugreek m which is smaller than the film thickness expected for hot drops ($h_0 \simeq 50$ \textmugreek m), but sufficient to enable levitation. Moreover, Eq. \ref{eqH1} predicts $h_{N_2} \propto R^{1/3}$, which we verified: indeed, Eq. \ref{eqTermV}, predicts that $V_A^*$, proportional to $h = h_{N_2}$ here, should also vary as $R^{1/3}$. The velocity $V_{A}^{*}$ of liquid nitrogen drops was measured for varying radii $R$ (inset of Figure \ref{figure5}b): the dotted line shows our model with $V_A^* = 1.7 \frac{\rho g}{\eta_v} \Delta h \, h_{N_2}$, which fits reasonably well with our data, with a prefactor close to 1.

\smallskip

The same mechanism that enables non-coalescence of liquid nitrogen drops also causes the persistence of levitation of initially hot droplets, long after they freeze to liquid nitrogen temperatures. Indeed, from Eq. \ref{eqTermV} and \ref{eqH}, one would expect the drops to sink at the end of the deceleration phase, when the film thickness $h$ diminishes to zero. However, the bath residual evaporation - as described earlier - generates a constant vapor flux that adds up to the Leidenfrost flux. Even if this additional vapor flux is negligible in the first seconds (it is initially 100 times smaller), it becomes of critical importance as the levitating drops cool down. Indeed, it generates a 10\,\textmugreek m thick vapor film (as estimated from Eq. \ref{eqH1}), which is sufficient to maintain in levitation droplets sufficiently light and smooth. 

\subsection*{Full model and discussion} To also model the dynamics of drops after they completely cool down (as in Fig. \ref{figure2}a), we now consider the influence of the residual vapor flux on the film thickness $h$. The calculation is provided in the SI Appendix: by adding the two fluxes, the film thickness $h(t)$ is found to be the solution of a polynomial equation: $h(t)^4 = \left ( h_0(1-t/\tau)\right )^4 + h_{N_2}^3h(t)$, that can be solved for any time $t$. We solved Eq \ref{NewtonsLaw} numerically by taking this last element into account: the continuous lines plotted in Figures \ref{figure2} and \ref{figure4} show the velocity profiles predicted by the model, with fitting parameters $\alpha$\,=\,$\beta$\,=\,15 and $\Delta h$\,=\,1.45\,\textmugreek m; $\tau$  and $h_{N_2}$ are calculated from Eq. \ref{eqH} and \ref{eqH1}, respectively. As seen in Figure \ref{figure2}a, the model matches all three phases of the drop movement. It also nicely reproduces the drop dynamics for varying initial conditions (Fig. \ref{figure2}b), liquid nature (Fig. \ref{figure2}c) and drop radii (Fig. \ref{figure4}a) without any change in the fitting parameters. This model, although simplified (it does not consider the variation of liquids properties as the droplets cool down, as well as the freezing dynamics) accounts convincingly for the details of the evolution of velocity amplitudes.

Remarkably, both experiments and numerical simulations are consistent with a stable symmetry breaking $\Delta h$, which remains constant during the drop's lifetime (even if the vapor flux diminishes by a factor 100), and which does not vary with the drop size. Even if we cannot fully explain the exceptional stability of the asymmetric state, we can provide clues on its origin. In particular, the consistent motion of non-deformable objects (frozen drops or polyethylene marbles, as in Movie S2) indicates that $\Delta h$ most certainly originates from an asymmetric deformation of the liquid nitrogen interface. What could then cause the surface of the bath to deform? A hypothesis is that the symmetry breaking is generated by an instability of the morphology of the vapor film itself, which is very different from that of classical Leidenfrost drops over a flat rigid substrate \cite{Maquet:2016, Wong:2017}. In particular, a recent theoretical study \cite{vanLimbeek:2018} shows that the film exhibits localized oscillations at the neck, which can develop within the whole film for drops smaller than the capillary length. We surmise that these oscillations may be unstable, which would trigger a symmetry breaking when they are very slightly disturbed.

\subsection*{Conclusion} We demonstrate that drops deposited on a cold bath are naturally self-propelled, without external forcing. The complexity of drop dynamics results from the combination of three elements: i) a stable symmetry breaking (associated with a variation $\Delta h$ of the film thickness) which causes self-propulsion ii) the thinning of the vapor film under the drops - due to their cooling -  that increases the friction and is responsible for their deceleration and, finally, iii) the residual evaporation of the bath, which can cause persistent levitation long after drops freeze to the bath temperature. 

An interesting parallel can be drawn with the very recent paper of A. Bouillant \textit{et al.} \cite{Bouillant:2018} who showed that small drops can also exhibit spontaneous self-propulsion on flat solids. While, on solids, motion is induced by a symmetry breaking in the internal flow of the droplets, on a bath, it is most surely generated by an instability happening at the liquid nitrogen interface. This difference fundamentally affects the behavior of the levitating objects: first, solid marbles can self-propel here and they glide without rotation. In addition, the propelling force switches direction and instantly reappears after the drops have been reflected from a wall. This can be used to control droplets trajectories with very fine precision, by confining them between two walls. We can finally note that spontaneous motion is not solely limited to cryogenic baths: liquid nitrogen drops can also self-propel on a ethanol bath (as in Movie S8). This might increase the scope of such a study to ambient temperature situations.

\matmethods{

 \textbf{Homemade cryostat}. The cryostat is a box of expanded polystyrene, with dimensions 30 x 30 x 25 cm and 4 cm thick walls. Inside is placed a sacrificial bath (a beaker with diameter 19 cm filled with 5 cm of liquid nitrogen). At its center, another beaker, with diameter $D$ = 7.6 cm is placed on a copper disk and filled with 10 cm of liquid nitrogen. The cryostat is closed by a polystyrene lid, which is removed for each experiment and then replaced.
 
\textbf{Drops tracking}. A home-made Python algorithm is used: it automatically extracts the (x,y) position of the drop center from an initial frame with known drop position and size. Bilateral filtering and median-estimated background subtraction are first applied. Then, at each step the drop position is estimated (from the previously tracked position and speed) and the image is cropped around it. A gaussian-blurred circle is drawn separately, and its center and radius are optimized through brute-force search to minimize its mean-squared error with the cropped image. This gives the drop location and radius with pixel precision.

\textbf{Numerical method}. The numerical simulation is based on a finite element method of the incompressible 2d Cartesian Navier-Stokes equations with sharp interfaces aligned with the mesh (see SI Appendix for more details). The two-dimensional simulation domain has a size of \SI{77}{\milli\meter} x \SI{45}{\milli\meter}. The liquid surface is placed at a height of \SI{20}{\milli\meter}, with a contact angle of \SI{20}{\degree} with respect to the walls. The implementation is done using the framework \textsc{oomph-lib}.\cite{Heil:2006}.

}

\showmatmethods % Display the Materials and Methods section

\acknow{The authors warmly thank Dominic Vella for insightful comments on the model, Corentin Tregouet for his initial and fruitful theoretical input and Guillaume Lajoinie for carefully reading the manuscript.}
\showacknow % Display the acknowledgments section

% \pnasbreak splits and balances the columns before the references.
% If you see unexpected formatting errors, try commenting out this line
% as it can run into problems with floats and footnotes on the final page.
%\pnasbreak

% Bibliography
\bibliography{pnas-sample}

\bibliographystyle{unsrt}
%\bibliography{biblio} 

\end{document}